# Reflector-Free, Highly Confined Love-Like SAWs Enabled by a Phononic Metasurface for Real-Time Monitoring of Cell Dynamics

Jessica Monaldi[1,2], Francis Kosior[1], Julio A. Iglesias Martínez[1], Laurent Badie[1], Halima Alem[1,3], Frédéric Sarry[1,4,5], and Mourad Oudich[1,3,*]

[1]*Université de Lorraine, CNRS, IJL, F-54000 Nancy, France*
[2]*Centre national d'études spatiales (CNES), 2 Place Maurice Quentin, 75039, Paris, France*
[3]*Institut Universitaire de France (IUF), F-75231, Paris, France*
[4]*Institut Interdisciplinaire d'Innovation Technologique (3IT), Université de Sherbrooke, 3000 Boulevard de l'université, Sherbrooke, J1K OA5 Québec, Canada*
[5]*Laboratoire Nanotechnologies Nanosystèmes (LN2)-IRL3463, CNRS, Université de Sherbrooke, INSA Lyon, École Centrale de Lyon, Université Grenoble Alpes, Sherbrooke, J1K 0A5 Québec, Canada*

*Corresponding author: mourad.oudich@univ-lorraine.fr

## Abstract

Surface acoustic wave (SAW) devices are widely used in sensing and biosensing but generally suffer from strong attenuation in liquid environments. Conventional approaches rely on reflectors to reduce these losses, yet these components remain difficult to optimize: limited device miniaturization, and increase fabrication complexity. Here, we introduce an innovative design strategy that integrates a phononic metasurface with tailored electromechanical properties of the substrate to generate a type of shear-horizontal (SH) surface resonance modes that exhibit strong lateral confinement and extremely low radiation into both the substrate bulk and the free surface, eliminating the need for reflectors. This approach enables highly tailorable surface acoustic resonances with distinctive enhanced dynamic strain-energy confinement leading to significantly higher quality factors than conventional SAW devices, particularly in water-loaded conditions. We show the fabrication and experimental validation of the proposed phononic metasurface-based SAW resonator and showcase its biosensing capabilities through real-time monitoring of cellular death.

## Introduction

For decades, surface acoustic wave (SAW) devices have been extensively developed for a wide range of applications, including telecommunications[1–3], sensing[4] and bio-sensing[5,6], acousto-fluidics[7–10], and recently, quantum optomechanics[11,12] and condensed matter physics[13,14]. In SAW sensing and biosensing in liquid environments, efforts have primarily focused on generating shear-horizontal SAWs (SH-SAWs), a type of surface waves characterized by in-plane solid-particle displacement perpendicular to the direction of propagation. Unlike Rayleigh-type SAWs, whose displacement is elliptically polarized within the sagittal plane and therefore radiates substantial acoustic energy into liquids, SH-SAWs exhibit minimal in-liquid radiation loss, making them particularly well suited for sensing surface perturbations in aqueous environments[6]. Consequently, most device-design strategies rely on either pure SH-SAWs or Love waves, the latter being SH-type SAWs produced by adding a low-velocity guiding layer on the substrate[6]. Several studies employed SH and Love waves in delay-line configurations to monitor biological processes[15–19] such as cell spreading and adhesion[18,19]. Despite these advances, the limited acoustic confinement and moderate quality factors achievable in liquid environments continue to restrict the frequency resolution and sensing performance of these devices, limiting their ability to monitor dynamic biological processes with the speed and precision achieved by state-of-the-art optical techniques.

Optical techniques, such as surface plasmon resonance, primarily detect changes in the refractive index of the surrounding medium, making them highly sensitive to molecular binding events. In contrast, SAW sensors respond to mechanical perturbations at the sensing interface, including mass loading, elasticity, viscosity, and viscoelastic changes, making them particularly well suited for monitoring dynamic biomechanical processes. The growing interest in SAW-based biosensors is further driven by their low cost, compact footprint, compatibility with lab-on-chip technologies, potential for wireless operation, and low power consumption.

Nevertheless, achieving efficient confinement and manipulation of SH-SAWs remains a significant challenge. Conventional SAW resonators rely on Bragg reflectors to form resonant cavities through acoustic reflection. Although highly effective, this approach increases device complexity and provides only indirect confinement of the acoustic field, making the achievable quality factor (Q-factor) sensitive to propagation losses, reflector imperfections, and fabrication tolerances. Moreover, in liquid environments, intrinsic loss mechanisms such as viscous loading and acoustic dissipation further degrade the Q-factor, limiting the performance of resonator-based biosensors.

A promising yet largely unexplored pathway to overcome these limitations lies in engineering SAW generation directly at the material surface. By engineering the substrate surface, it becomes possible to tailor SH-modes for specific sensing environments, potentially enabling substantial gains in sensitivity. In this context, integrating artificial materials such as phononic crystals (PnCs) and acoustic metamaterials at the substrate surface is highly desirable, as these structures could achieve strong mode confinement and waveguiding without the need for reflectors. Despite their promise, such approaches remain unexplored for SAWs, particularly for SH-type modes designed for in-liquid biosensing.

For nearly three decades, the field of PnCs and acoustic metamaterials has flourished, driven by the development of diverse artificial structures operating across a broad range of frequencies. These systems target functionalities spanning seismic shielding[20,21], vibration mitigation[22], ultrasound control[23,24], and even quantum optomechanics[25] and phononic circuit[26,27]. In the context of SAW manipulation in microelectromechanical systems, PnCs have been predominantly explored for enhanced filtering[28–32] and waveguiding[33,34] through the creation of phononic bandgaps. Early studies introduced PnC lattices of holes drilled into a piezoelectric substrate to create SAW bandgaps for filtering purposes[28,29]. Subsequently, acoustic metamaterials consisting of lattices of pillars on the substrate surface attracted considerable attention for wave filtering[30–32], waveguiding[34], and more recently sensing[35,32], leveraging the local resonances of the pillars excited by SAWs. For instance, Bonhomme *et al.*[32] designed a Love-SAW sensor incorporating a lattice of subwavelength resonant pillars to achieve temperature and microbead concentration sensing in liquid environment. In the context of SAW resonators, Gao et al.[36] developed a device based on a PnC formed by an array of high–aspect-ratio ridges capable of operating in liquid conditions. In their experiments, they achieved an in-liquid Q-factor of roughly 50, about fifteen times higher than that of a conventional Rayleigh-wave resonator at a similar operating frequency. This resonator, however, relied on reflectors to enhance wave-energy confinement within the ridge array, and the bio-sensing functionality was not demonstrated at that stage. These advances highlight the strong potential of PnC and metamaterial-integrated platforms for enhanced SAW sensing and biosensing, particularly in terms of improving sensitivity and Q-factor in liquid environments. Yet, most SAW sensing strategies still rely on delay-line configurations that detect perturbations along a free propagation path, often supplemented with reflectors to boost transmission. Even the limited number of SAW-resonator-based studies typically incorporate reflectors to reinforce the signal and suppress energy leakage into undesired regions. This reliance on reflectors, however, complicates device architecture and makes it difficult to optimize sensitivity and Q-factor, especially in aqueous environments.

In this work, we introduce a new design paradigm for achieving highly confined SH Love-like surface resonances without the need for Bragg reflector gratings. Conventional Rayleigh, SH, and Love surface acoustic waves are primarily governed by the intrinsic elastic properties and crystallographic orientation of the substrate. By introducing a phononic metasurface, however, the acoustic dispersion can be engineered through the geometry of the periodic structure, providing an additional degree of freedom to tailor wave confinement, propagation, and energy localization beyond those dictated by the substrate alone. Here, we combine a carefully engineered phononic metasurface with an ST-90°X quartz substrate, whose crystallographic orientation exhibits vanishing electromechanical coupling for propagating SH-SAWs[37]. The phononic metasurface, composed of periodically arranged heavy-metal ridges, supports a confined Love-like SH mode below the sound line, resulting in strong suppression of bulk radiation and rapid evanescent decay outside the patterned region. Consequently, acoustic confinement is achieved without the use of additional reflector gratings. Throughout this work, the term ‘reflector-free’ refers specifically to the absence of such external reflector gratings; the necessary resonant feedback is instead provided by the finite phononic metasurface itself. Rather than introducing a new class of acoustic waves, our approach establishes a new strategy for engineering the dispersion and confinement of conventional SH surface waves through phononic metasurface design. This approach results in sharp resonances with high Q-factors in both air and liquid environments while providing a simple and compact device architecture. To the best of our knowledge, it exhibits one of the highest reported Q-factors for reflector-free SH-SAW resonators operating in liquid. Finally, the practical potential of the proposed platform for biosensing is demonstrated through real-time monitoring of cell death, highlighting its suitability for high-performance sensing in biologically relevant liquid environments.

## Result and discussion

### 1. The phononic SAW resonator

The phononic metasurface consists of a one-dimensional array of gold ridges deposited on an ST-cut quartz substrate rotated by 90° with respect to the X-axis (Euler angles: (0°, 132.75°, 90°)) (**Fig. 1(a)**). This crystallographic orientation was selected because it exhibits zero electromechanical coupling for propagating SAW on the free (unpatterned) surface, thereby preventing their direct piezoelectric excitation and propagation along the substrate[37]. For wave propagation along the X direction, only the fundamental shear-horizontal (SH) bulk mode exhibits significant electromechanical coupling. Consequently, while SH bulk waves and SH-type pseudo-SAWs can be excited piezoelectrically, conventional Rayleigh waves cannot, and true SH-SAWs do not naturally exist on the free surface for this crystal orientation[37]. This apparent limitation is overcome by introducing the phononic metasurface. The periodic array of heavy gold ridges modifies the local acoustic dispersion through mass loading, reducing the phase velocity of the SH-type pseudo-SAW. Once this velocity falls below the bulk SH-wave sound line, radiation into the substrate is strongly suppressed, and the pseudo-SAW evolves into a strongly confined Love-like SH surface mode. The phononic metasurface therefore enables the existence of a localized SH resonance that is not supported by the unpatterned substrate alone. In the proposed architecture, the metallic phononic metasurface plays a dual role: it engineers the acoustic dispersion to create the confined Love-like mode while simultaneously acting as the interdigital transducer for its electromechanical excitation. Because the acoustic energy is confined within the finite phononic metasurface itself, no additional Bragg reflector gratings are required to realize the resonator.

Finite element simulations were conducted using the commercial software Comsol Multiphysics (See Methods). The ridges length is $l_{\mathrm{Au}} = 2.5\ \mu m$, with periodicity of $p = 5\ \mu m$, while the height is denoted $h_{\mathrm{Au}}$ (**Fig. 1(a)**). The calculated band structure of the infinite array (**Fig. 1(b)**, left) with $h_{\mathrm{Au}} = 500\ nm$ reveals three surface modes below the sound line: two Rayleigh-like modes ($R_1$ and $R_2$) and one SH or Love-like mode (L). The Rayleigh-like modes exhibit no electromechanical coupling, as indicated by their negligible electrical potential (**Fig. 1(b)**, right). In contrast, the L mode shows a non-zero electrical potential, indicating that it can be excited electromechanically. For this particular mode, the electromechanical coupling coefficient is numerically evaluated to 0.31 %. The appearance of these sub–sound-line bands is attributed to the heavy gold ridges, which lower the dispersion curves with the mass effect, and creates particular dispersive surface modes. Among these modes, only the L-SAW can be driven electrically. This demonstrates that, even when starting from a substrate orientation with zero intrinsic electromechanical coupling for the generation of SAW, an appropriately designed phononic metasurface can unlock surface acoustic modes below the sound line. Two effects are exploited here: (i) the mass of the gold ridges, which pulls the dispersion below the sound line, and (ii) the ridges acting as an effective guiding layer that supports a Love-like SAW.

For a finite number of ridges, this configuration enables a resonant stationary mode with very weak leakage into the bulk and into the unpatterned free surface of the ST quartz substrate, leading to strong acoustic confinement within the metasurface. The wave field becomes strongly evanescent in the free surface. Additionally, the metallic ridges serve as electrodes, i.e., interdigital transducers (IDTs), allowing direct excitation of the L surface mode only within the phononic metasurface. To demonstrate this, we performed numerical simulations of a finite array ridges with electromechanical excitation using a harmonic voltage and ground connections, as in classical IDT-based SAW generation. The resulting device is a reflector-free SAW resonator supporting a highly confined Love-like mode without leakage into the bulk nor into the unpatterned free surface (**Fig.2 (c)** and **(d)**). **Figure 2(a)** shows the impedance magnitude as a function of frequency for different numbers of ridge pairs (each pair consisting of one grounded ridge and one excited ridge). A clear resonance peak corresponding to the L-SAW mode is observed. As the number of pairs increases from 10 to 45, the resonance frequency converges toward that of the infinite array, around 253 MHz (**Fig. 2(b)**). The wavelength is set by the lattice periodicity, $\lambda = 2p$. **Figures 2(c)** and **2(d)** show the displacement component $u_y$ at resonance, illustrating the strong localization of the L mode within the phononic array, with no propagation into the bulk nor into the unpatterned quartz surface. This confirms that external reflectors are unnecessary for this phononic resonator case, as the surface wave becomes strongly evanescent on the free surface. Moreover, the displacement field tend to be uniform along the ridges for a higher number of ridges as it can be deduced for the case of 90 ridges (45 pairs) (**Fig. 2(d)**).

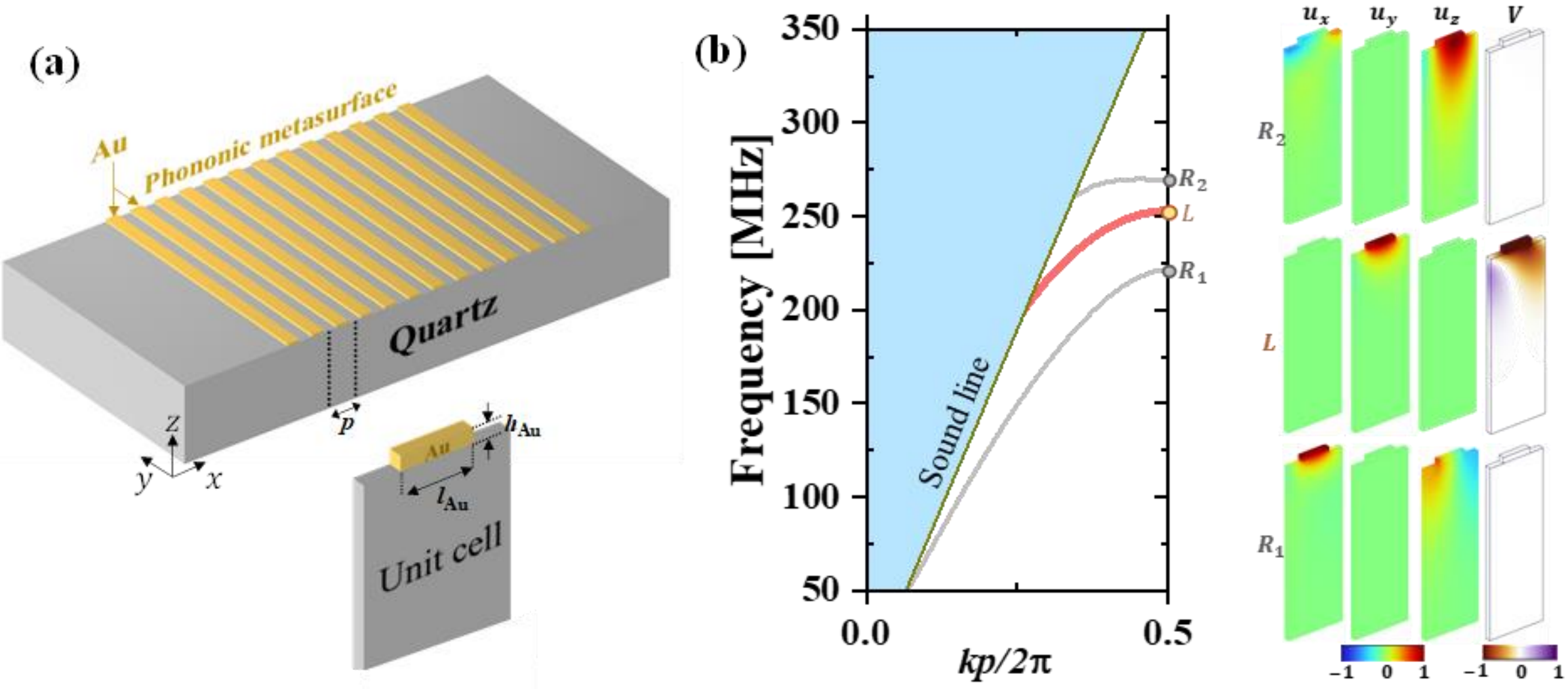

**FIG 1**. (a) The phononic metasurface made of heavy gold ridges in quartz ST + 90° X- rotated substrate, and (b) (left panel) its associated band structure for the case where the height and the length of the ridges are $h = 500\ nm$, and $l = 2.5\ \mu m$ respectively, while the period is $p = 5\ \mu m$. The right panel of (b) present the displacement field components $u_x$, $u_y$, and $u_z$, and the electrical potential V for the three depicted surface modes $R_1$, $R_2$ and L in the band structure at the X point. Only the L mode has non-zero electrical potential, hence it can be excited electromechanically.

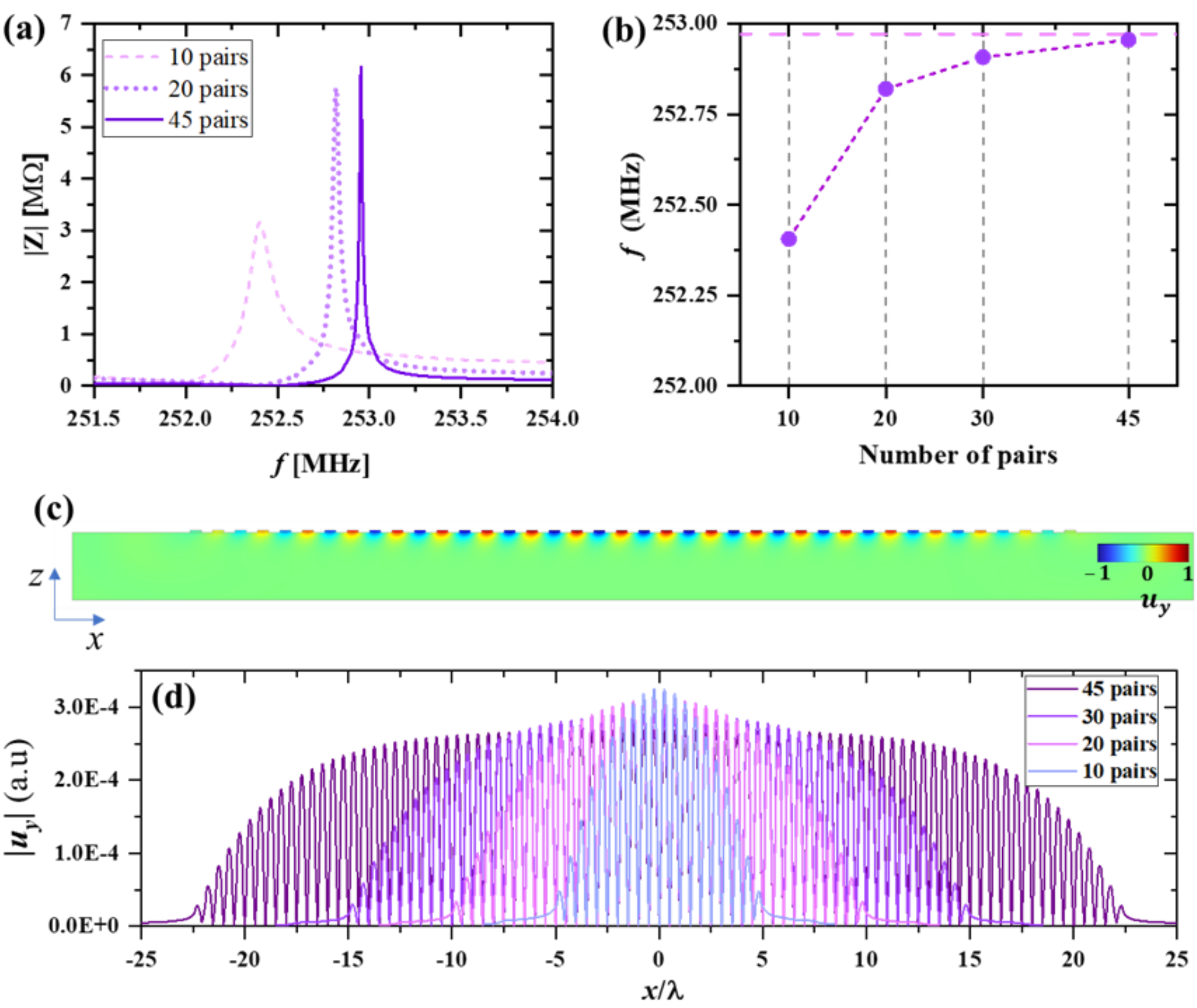

**FIG. 2**. (a) Calculated absolute value of the impedance for the finite lattice of the phononic metasurface for different numbers of pairs, showing the resonance associated with the excitation of the Love-like surface mode. (b) Evolution of the L-mode frequency as function of the number of the IDT pairs. The dashed line is for the case of infinite array of ridges. (c) Displacement field component $u_y$ at the resonance for the case of 20 pairs. (d) amplitude of the displacement component $u_y$ as function of the position along the surface for different numbers of IDTs pairs. Here $\lambda = 2p$.

## 2. Device fabrication and characterization in air

Using clean room facilities (see Methods), the phononic metasurface was fabricated and characterized. In all the devices, we deposited an array of 90 ridges of gold on the quartz ST substrate (**Fig. 3(a)**). Several devices were fabricated with different values of thickness of gold depositions for the ridges $h_{\mathrm{Au}} = 150nm$, $250\ nm$, and $500\ nm$; for an operating wavelength of $20\mu m$. The electrical impedance of the phononic resonator was measured and an example of result for three devices is presented in **Fig. 3(b)** corresponding to the three values of the ridge thickness. The devices manifest a Fano resonance type peak corresponding to the excitation of the confined L-SAW mode within the phononic ridges. As the thickness of the ridges is increased from $150nm$ to $500nm$, the resonance mode is pushed down to lower frequencies thanks to the increased mass effect of the ridges: from 174.7 MHz for $h_{\mathrm{Au}} = 500nm$ to 213.6 MHz for $h_{\mathrm{Au}} = 150nm$. Good agreement is found between numerical prediction and the measured resonance frequencies of the devices for the three different values of $h_{\mathrm{Au}}$ (**Fig. 3(c)**). The mass sensitivity of the device can be evaluated (see Method), and the simulation value is -83.7Hz.cm$^2$/ng while the experimental one is -67.35 Hz.cm$^2$/ng which is considered high in literature[36].

We also evaluated the Q-factors of these resonances for the three cases of ridge thicknesses $h_{\mathrm{Au}} = 150nm$, $250\ nm$, and $500\ nm$, while fabricating three devices for each case, and results are presented with error bars in **Fig. 3(d)**. The figure shows that the Q-factor can be higher than 1700 for the case of $h_{\mathrm{Au}} = 250\ nm$ which is very promising for high-precision sensing. For this same thickness, devices were fabricated with different wavelengths $\lambda = 10\mu m$, $16\mu m$, and $20\mu m$ (periodicity of ridges of $p = 5\mu m$, $8\mu m$, and $10\mu m$), and very good agreement is found between numerical prediction and measurements for the resonance frequency (**Fig. 3(e)**). The frequency decreases almost linearly with the operating wavelength as expected.

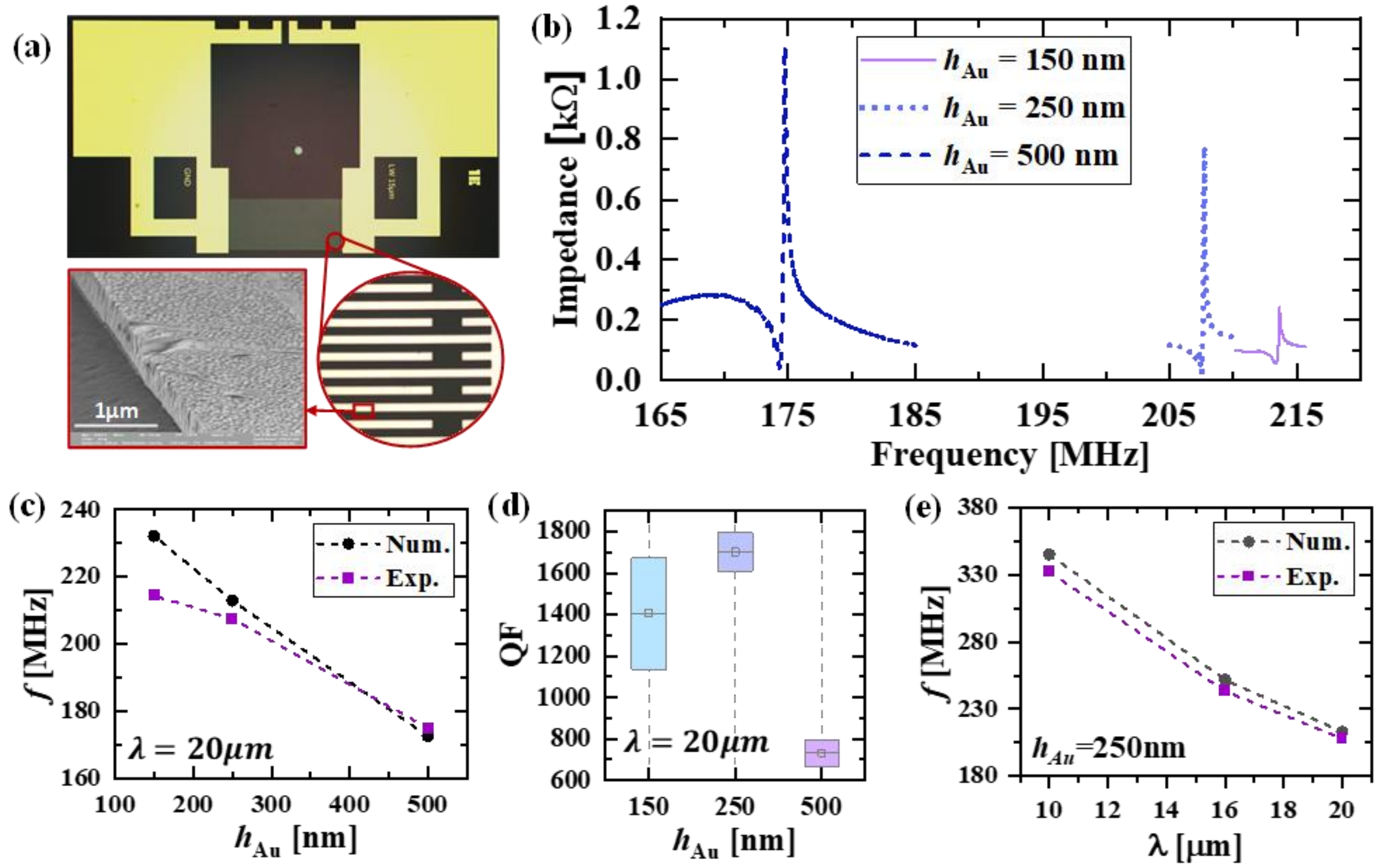

**FIG 3**. (a) Optical and SEM image of the fabricated phononic metasurface resonator made of heavy gold ridges in Quartz ST. (b) Measured impedance amplitude for three devices with different ridge thickness $h_{\text{Au}} = 150nm, 250\ nm, 500\ nm$, for $\lambda = 20\mu m$. (c) Numerical and experimental measurement of the resonance frequency of the L-SAW as function of the thickness of the ridges for a fixed value of the wavelength $\lambda = 20\mu m$. (d) Q-factor (QF) of the L-SAW resonance mode as function of the ridge thickness for three fabricated devices for each thickness. Colored rectangles are error bars. (e) Numerical and experimental values of the resonance frequency of the L-SAW as function of the operating wavelength for $h_{\text{Au}} = 250nm$.

We would like to point out that the device operates in the linear regime under the applied excitation conditions, with no observable nonlinear contributions. To verify this, impedance measurements were performed at different input power levels. The resonance response remains unchanged over the investigated range, indicating that nonlinear effects are negligible. Significant nonlinear wave generation would induce energy transfer from the fundamental mode to higher-order modes, resulting in measurable changes in the impedance response, which are not observed here.

## 3. Characterization in aqueous media and real-time monitoring of cells death

After validating the operation of the phononic metasurface–based SAW resonator in air, we optimized the device for sensing in aqueous environments. In the following studies, we focused on devices with a ridge thickness of $h_{\text{Au}} = 250\ nm$ as this configuration exhibits the highest Q-factor (**Fig. 3(d)**). To prepare the samples for liquid-phase sensing and to prevent any direct contact between the metal and the fluid, the ridges were covered with a 300 nm-thick silicon oxide layer (See Methods). This layer serves both as electrical insulation to prevent short-circuiting and as a means to reduce signal loss in aqueous environments. **Figure 4(a)** shows a SEM cross-sectional image of the device, revealing the silicon oxide ($SiO_2$) layer covering the gold ridges and their thicknesses. Three sets of samples were fabricated with this $SiO_2$/Au bilayer configuration. Experimental measurements show that the addition of the silicon oxide layer induces a shift in the resonance frequency from 207.7 MHz to 205.9 MHz (less than 1% variation) due to the slight added mass loading effect (**Fig. 4(b)**), while the impedance amplitude remains well preserved at the resonance. The resonance frequency exhibits only minor variations upon the addition of the silicon oxide layer for all three devices with operating wavelengths of 10 μm, 16 μm, and 20 μm, as shown in **Fig. 4(c)**. We also observed a slight increase in the Q-factor in air, from approximately 1650 before passivation to 1700 after passivation with the $SiO_2$ layer. This modest improvement can be attributed to the increased effective stiffness introduced by the $SiO_2$ layer, which slightly enhances the confinement of the SH mode within the sensing region. The Q-factor of the three devices was characterized, revealing that the device with an operating wavelength of 20 μm exhibits the highest Q-factor, reaching an average value around 1300 in air (**Fig. 4(d)**).

A microfluidic chamber was then integrated on top of the SAW resonator (see Methods) (**Fig. 4(e)**) to enable characterization in aqueous environments and to assess the biosensing performance of the phononic metasurface SAW resonator. Initial experiments were conducted by filling the microfluidic chamber with water. Despite the viscous losses introduced by the liquid, the measured electrical impedance amplitude still exhibits well-defined and clearly identifiable resonance peaks corresponding to the phononic mode for all three operating wavelengths (10 μm, 16 μm, and 20 μm), as shown in **Fig. 4(f)**. The extracted Q-factors in water remain remarkably high, with average values exceeding 135 for all three wavelengths and reaching values as high as 225 and 250 for devices with $\lambda = 16\mu m$, and $\lambda = 20\mu m$ , respectively (**Fig. 4(g)**). Although water reduces the Q-factor because of viscosity which leads to energy dissipation, these values are at least 5 times higher than what is reported in literature of SAW resonator with reflectors in aqueous environment[36]. Such sharp underwater resonances translate directly into a significant enhancement in sensing precision, while simultaneously offering a much simpler design approach and reduced fabrication constraints.

Numerous conventional Love-wave devices exhibit good Q-factors under appropriate conditions. A literature comparison is given in **Table 1**, specifically with SAW devices reported operating in liquid environments, where viscous damping significantly reduces the Q-factor. We note, however, that a fully quantitative comparison is challenging because several published studies do not report key parameters such as Q-factor or mass sensitivity. Furthermore, the definition of sensitivity varies considerably across the literature, being expressed, for example, in terms of frequency shift, phase shift, or analyte concentration depending on the sensing methodology and target application. Nevertheless, **Table 1** provides an overview of the device architectures, operating principles, reflector grating usage, Q-factors, sensing metrics, and biological functionalities reported in the literature. This comparison shows that the proposed phononic metasurface platform achieves highly competitive performance, combining one of the highest reported Q-factors in liquid operation with very good sensitivity while introducing a fundamentally different reflector-free operating principle.

We would like to point out that in air, the Q-factor is primarily limited by intrinsic material damping in the quartz substrate and metallic electrodes, fabrication imperfections, and weak radiation losses associated with the finite extent of the phononic metasurface. Although the engineered SH mode is strongly confined within the metasurface, its evanescent field extends slightly beyond the patterned region, leading to a small residual leakage into the surrounding substrate. In liquid, viscous loading becomes the dominant dissipation mechanism, resulting in additional acoustic energy losses and a corresponding reduction of the Q-factor. Despite these losses, the proposed resonator maintains a high Q-factor in liquid, highlighting the effectiveness of the phononic metasurface in preserving strong acoustic confinement under biologically relevant operating conditions.

As a preliminary assessment of the biosensing performance of our device, we conducted real-time measurements to monitor cell death. For these experiments, we used the phononic device operating at a wavelength of 20 μm, which exhibits the highest in-liquid Q-factor, with an average value of 250 (**Fig. 4(g)**). After standard preparation procedures (see Methods), approximately 100k MeWo cells were seeded onto the device within a microfluidic chamber of surface area 35 $mm^2$. First, we compare the response in different liquid environments and evaluate the stability of the resonator by performing measurements using water, cell-culture medium, and a 15% ethanol (EtOH) aqueous solution. The resonance frequency was monitored continuously over a period of 10 min, and the results are presented in **Fig. 4(h)**. The frequency shifts induced by water and the cell-culture medium are nearly identical and remain stable throughout the measurement, demonstrating that the resonator exhibits very good stability under both conditions. In contrast, the 15% EtOH solution produces a significantly larger frequency shift, which remains clearly distinguishable from the response obtained with the culture medium. A slight drift is observed over time for the EtOH solution, which we attribute to the gradual evaporation of ethanol due to its volatile nature.

For the cell death test, two sets of experiments were performed. First, a reference measurement was carried out with cells immersed in their culture medium, during which the resonance frequency shift and phase were monitored continuously for 10 min. Subsequently, the culture medium was replaced by a 15% ethanol (EtOH) solution to rapidly induce cell death, and the system response was continuously recorded for 10 min under identical acquisition conditions (see Methods). **Figure 4(i)** presents the temporal evolution of the resonance frequency shift and phase for both experimental conditions. With the cells in their culture medium (square symbols), no significant frequency shift (filled squares) or phase variation (open squares) is observed over 10 min. By contrast, upon introduction of the EtOH solution in the presence of cells, a pronounced change in both the resonance frequency (filled circles) and phase (open circles) is detected, following an exponential evolution until a steady state is reached. This response is attributed to changes in the local viscosity near the device surface as cell death causes the cells to detach from the surface. These results clearly demonstrate real-time monitoring of cell death, with a resonance frequency variation exceeding 0.15 MHz and achieved with high precision.

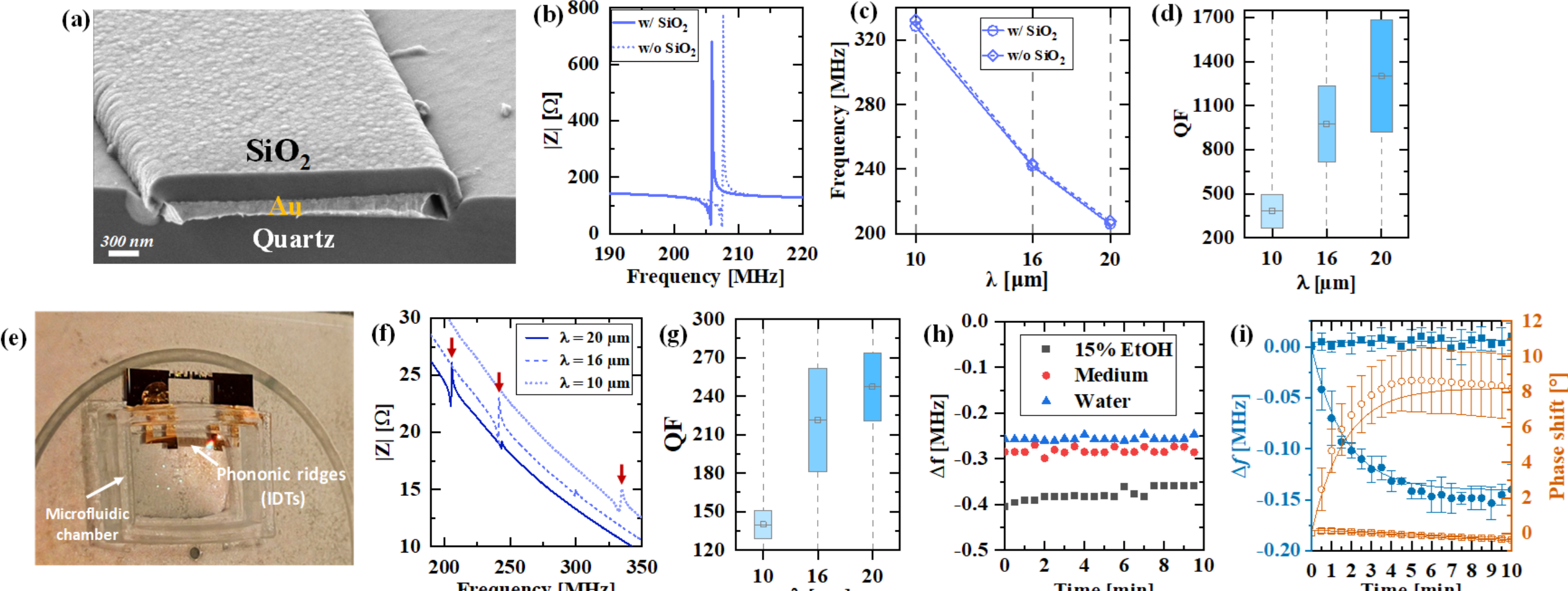

**Fig. 4.** (a) SEM image of the fabricated gold phononic micro-ridge with the deposited silicon oxide layer. (b) Impedance amplitude at resonance for devices with λ = 20 µm, comparing samples with and without the silicon oxide layer. (c) Resonance frequencies of devices with λ = 10 µm, 16 µm, and 20 µm, with and without the silicon oxide layer. (d) Q-factor (QF) for the three operating wavelengths, including error bars and average values; multiple samples were fabricated for each wavelength with the silicon oxide layer. (e) The phononic SAW device with a mounted microfluidic chamber. (f) Measured impedance amplitude in water covering the phononic ridges inside the microfluidic chamber; arrows indicate the resonance peak for each device. (g) Q-factor of the devices measured under water. (h) Comparison of different liquids (water, cell-culture medium, and solution with 15% EtOH) during a 10 min-window observation. (i) Frequency shift and phase change at resonance as a function of time during cell-death monitoring. A total of 100k MeWo cells were used. Filled squares correspond to cells in standard culture medium, while open circles correspond to medium containing solution of 15% ethanol (EtOH).

| Study | Device architecture | Substrate | Reflectors | Operating frequency | SAW type | QF in air | QF in liquid | Mass sensitivity | Biosensing functionality |
|---|---|---|---|---|---|---|---|---|---|
| **This work** | **Single IDTs resonator** | **90° X ST-quartz** | **No** | **341; 248; 207 MHz** | **SH** | **1150** | **266** | **- 67.35 Hz·cm²/ng** | **Fast real-time monitoring of cells death** |
| Gao *et al.*[36] | High aspect ratio IDTs resonator | 128° Y-cut LN | Yes | 120; 130 MHz | Rayleigh | - | 50.4 | −12.92 Hz·cm²/ng | Underwater test, no biosensing |
| Bonhomme *et al.*[32] | Delay line IDTs: Gold GL: SU8 + SU8 pillars, | 36°-Y $LiNbO_3$ | No | 34.6 MHz | Love | 510 | Below 100 | −76.17 Hz·cm²/ng (deduced from microbead concentration) | Underwater microbeads concentration detection |
| Kustanovich et al. [38] | Resonator IDTs : Gold | Y cut $LiNbO_3$ | Yes | 185 MHz | SH | 300 | 250 | −90ppm (SLB to SLB/NPs), not a mass sensitivity | Biotin-modified SLB followed by neutravidin-coated nanoparticles (NPs) |

| Agostini *et al.*[39] | Two IDTs resonators | 128° Y-cut X-rotated LN | Yes | 1.285 GHz | Rayleigh | - | - | $-3.8\text{x}10^{3}$ Hz·cm²/ng | Streptavidin concentration |
|---|---|---|---|---|---|---|---|---|---|
| Schlensog *et al.*[15] | Delay-line, Gold IDTs and GL: $SiO_2$ | Quartz ST | Yes | 138-142 MHz | Love | - | - | 0.419 cm$^2$/ng (using phase shift) | Peptide detection |
| Saitakis *et al.*[16] | Delay-line IDTs: Gold GL: PMMA | Y-cut Quartz | - | 110-MHz | Love | - | - | - | Real time interaction monitoring of LG2 cells with surface-immobilized anti-HLA antibody |
| Li *et al.*[17] | Delay-line IDTs GL: SiO2 + gold nanoparticles functionalization | ST 90°-X quartz | No | 120 MHz | Love | - | - | - | Cancer-related biomarker carcinoembryonic antigen |
| Wu *et al.*[18] | Delay line IDTs: Gold GL: Gold + Parylene-C | 36°-YX LiTaO3 | - | 128 MHz | Love SAW | - | - | - | Tracking the adhesion process of tendon stem cells |
| Brugger *et al.*[19] | Delay line config. IDTs: Gold, 50nm; GL: SiO2, 150nm | 36°-YX $LiTaO_3$ | No | 208.4MHz | Love SAW | - | - | - | Wound healing assay by observing cell growth and quantifying cell detachment process |

**Table 1**. Comparison of representative SAW sensors operating in liquid environments, including device architecture, acoustic mode, Q-factor, sensitivity, and biological sensing demonstrations.

To confirm that the measured frequency variations are associated with changes in cell adhesion and viability, we performed time-resolved live/dead microscopy during the cell-detachment experiment (see the experimental protocol in the Methods section). The microscope was focused on the surface to monitor the evolution of the attached cell population throughout the measurement. The results are presented in (**Fig. 5**). When cells remain immersed in their culture medium (**Fig. 5(a)**), they remain attached to the substrate and the optical signal stays essentially unchanged over the entire observation period, confirming stable cell adhesion under physiological conditions. In contrast, when the culture medium is replaced by a 15% EtOH aqueous solution (**Fig. 5(b)**), the cells initially swell due to ethanol exposure and subsequently detach from the substrate as a consequence of ethanol-induced cell death. As detachment occurs, the cells progressively move out of the focal plane, providing direct optical evidence of the loss of surface adhesion. From the microscopy images, we quantified the fraction of cells remaining attached to the substrate as a function of time. The resulting curve, shown in **Fig. 5(c)**, exhibits an approximately exponential decrease in the number of attached cells after ethanol exposure (filled circles in the figure), while remaining nearly constant for the control condition (filled squares). This temporal evolution is consistent with the sensor response reported in **Fig. 4(i)**, providing independent optical validation that the measured frequency shift is associated with the progressive loss of cell adhesion during the cell-death process.

The observed frequency variation during ethanol exposure should not be interpreted solely as a consequence of a change in the local viscosity. Instead, the sensor response results from the modification of the acoustic boundary conditions at the surface induced by the dynamic evolution of the cell–substrate interface. The strongly confined SH acoustic mode within the phononic metasurface is highly sensitive to variations in the effective interfacial mechanical loading, including changes in cell adhesion, viscoelastic coupling, and mass distribution at the surface. Upon exposure to ethanol, cells undergo morphological changes followed by progressive loss of adhesion and detachment from the substrate, leading to a modification of the local acoustic interaction region and consequently to a measurable frequency shift. The frequency variation observed in our device is therefore attributed to the enhanced interaction between the highly confined acoustic field and the dynamically changing cell–surface interface, rather than to a simple change in the surrounding fluid properties

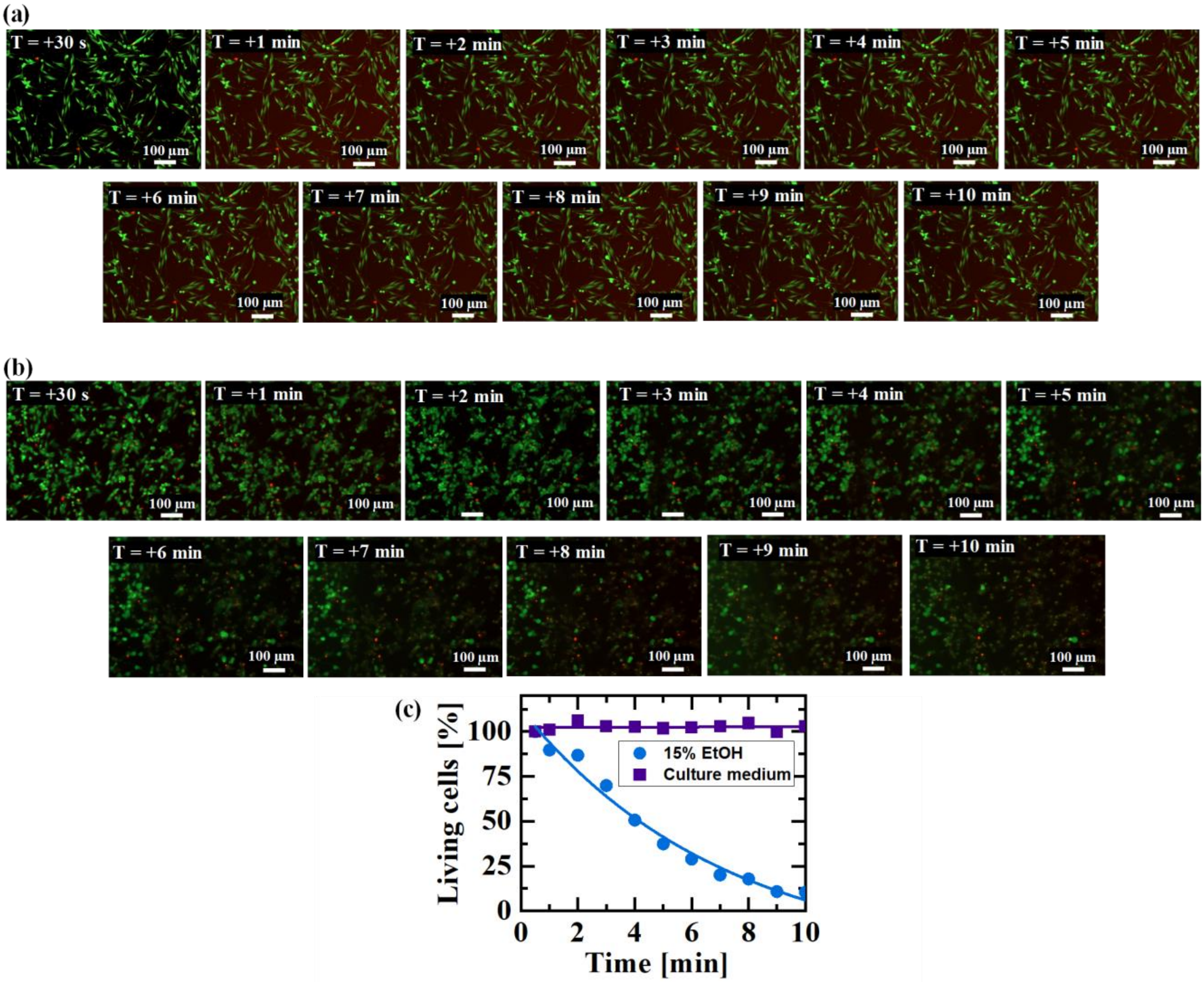

**Fig. 5**. Time-resolved live/dead assay of cells on the same sample over a 10 min period under (a) culture medium and (b) 15% EtOH solution. (c) Fraction of cells remaining attached to the substrate as a function of time, quantified from the microscopy images.

## Conclusion

The design of SAW sensing platforms that operate in liquid environments is challenging, as one of the major issues is the acoustic signal loss caused by energy radiation into the fluid and damping due to viscosity, especially when dealing with biological species. SH-type SAWs are widely used because they exhibit lower radiative energy loss in liquids, but their sensitivity and precision remain very limited and fail to compete with optical biosensing techniques. The use of reflectors reaches its limit in tightly confining the SAW within the sensing area and improving performance, while adding complexity to the design and fabrication. In our study, we incorporate an innovative design strategy that evolves the design of PnCs to shape the SAW field and create a category of dispersive acoustic waves that can enhance performance, while taking advantage of the substrate's electromechanical properties. By constructing a phononic metasurface made of heavy-metal ridges (gold) on an ST-90°X quartz substrate with zero electromechanical coupling for SH-type propagating SAWs, we create dispersive SH modes below the sound line, which eliminates both bulk and surface radiation outside the phononic ridges. Consequently, no reflectors are needed, providing simplicity in design and fabrication while enhancing the confinement of SAW energy within the phononic ridges. The resulting experimental sensitivity of the device reaches approximately -67.35 Hz·cm²/ng, while the measured in-liquid Q-factor of 266 significantly surpasses previously reported values, to the best of our knowledge. The phononic metasurface-based device shows high precision real-time monitoring of cell death at the surface, leading to high-performance sensing in biologically relevant liquid environments.

## Methods

**Numerical simulation**. Based on the design requirements and operating conditions of the device (biocompatibility, in-liquid operation, temperature stability, an operating wavelength comparable to cell dimensions, and compatibility with microfluidic integration), finite element analysis was conducted to guide and optimize the structure using the commercial software COMSOL Multiphysics 6.3. The Piezoelectricity module within Structural Mechanics was employed. Eigenfrequency analysis was used to compute the band structure, while frequency-domain simulations were carried out to excite the phononic resonance and evaluate both the impedance amplitude and the displacement field. The materials selected were: (1) ST–90°X quartz for the substrate, (2) gold (Au) for the phononic metasurface, chosen for its biocompatibility and high density relative to conventional electrode materials, and (3) a silicon oxide layer was included to prevent short-circuiting of the metallic ridges and to reduce signal loss in aqueous environments (water and cell-culture medium).

**Fabrication**. The phononic metasurface was patterned on an ST–90°X Quartz substrate using UV photolithography. A gold (Au) layer of the desired thickness was deposited via physical vapor deposition, and lift-off processing was used to define the phononic micro-ridges. A second photolithography step was then performed to form a 300 nm silicon oxide ($SiO_2$) passivation layer, deposited by magnetron sputtering to protect the metallic structures. Microfluidic wells were fabricated to isolate the biological samples. Poly(di-methyl-siloxane) (PDMS, 10:1) was cast onto molds and cured on a hot plate at 60 °C for at least 4 hours. The cured PDMS wells were then retrieved and bonded onto the phononic metasurface using oxygen-plasma treatment.

**Electrical characterization**. The performance of the resonator is evaluated using a vector network analyzer (VNA) which records the real and the imaginary components of the reflection coefficient $S_{11}$. From these data, the corresponding electrical impedance $Z_{11}$ is obtained. The Q-factor is then calculated from the magnitude of the impedance. The complex reflection coefficient is first reconstructed from its real and imaginary parts as

$$S_{11} = \Re\{S_{11}\} + i.\Im\{S_{11}\}$$

$Z_{11}$ is the characteristic impedance of the measurement system (typically 50 Ω),

$$Z_{11} = Z_0.\frac{1 + S_{11}}{1 - S_{11}}$$

where $Z_0$ is the characteristic impedance of the measurement system (50 Ω).

We extract the intrinsic Q-factor by fitting the experimental resonance spectra using a Fano line shape, which explicitly accounts for the resonance asymmetry:

$$A(f) = A_0 + B\,(q + \epsilon)^2/(1 + \epsilon^2)$$

with

$$\epsilon = 2\,(f - f_0)/\Gamma$$

where $f_0$ is the resonance frequency, Γ is the linewidth, $q$ is the Fano asymmetry parameter, $A_0$ the offset, and $B$ the scaling amplitude. The Q-factor is then obtained from,

$$Q = f_0/\Gamma$$

Regarding the evaluation of the mass sensitivity in practical biosensing applications, the added mass generally originates from adsorbed analytes on the sensor surface. In the present work, however, the sensitivity was intentionally evaluated by varying the electrode thickness in order to quantify the frequency shift induced by a well-controlled mass perturbation. This methodology has been adopted in several previous studies, including the work of Gao et al.[36], allowing a direct comparison with the state of the art while avoiding uncertainties associated with biological or deposited analytes. The mass sensitivity is calculated following Gao *et al*.[36] :

$$S = \frac{1}{\rho_{Au}} \frac{\Delta f}{\Delta h_{Au}}$$

where $\Delta f$ is the resonance frequency shift induced by a change in the ridge thickness $\Delta h_{Au}$ and $\rho_{Au}$ is the density of gold.

**Cell death tracking tests**. The cells used in the study are MeWo (HTB-65 datasheet from ATCC), BSL-1 (Biosafety in Microbiological and Biomedical Laboratories, from Centers for Disease Control and Prevention National Institutes of Health). The culturing medium used is Eagle's Minimum Essential Medium, with addition of 10% Fetal Bovine Serum (FBS), 1% Penicillin Streptomycin (Pen-Strep), 0.05% Amphotericin B.

Before the seeding of the cells on the devices, they are rinsed with 70% ethanol (EtOH) and dried to remove dust particles, then sterilized with UV light for 1 hour[40]. Having an available surface of around 35 mm², $10^5$ cells have been seeded in the wells, for a final concentration of around $2.5 \times 10^5$ cells per cm² [41,42], and left in an incubator with a temperature of 37°C and a $CO_2$ concentration of 5%. Due to the high concentration of the cell solution, cells are already in a confluence state after 24h, as proven by optical microscopy observation. So, the seeded devices are retrieved from the incubator for the cell death process tracking.

For the characterization with VNA, prior to each measurement, the microscope holding stage was cleaned with 70% ethanol (EtOH) and then brought to 23°C to ensure a uniform and reproducible temperature throughout the experiments, including measurements performed on different days. Once thermal equilibrium was reached, the lab-on-chip (LOC) device (SAW device including the microfluidic chamber) was placed on the stage and connected to the VNA probes. First, the response of cells immersed in their culture medium was monitored for 10 min, with spectral data recorded every 30 seconds. After this initial period, the culture medium was removed and replaced with 15% EtOH. The device response was then continuously monitored for an additional 10 min under the same acquisition conditions. The objective of this protocol is to detect resonance frequency shifts and phase changes associated with the onset of cell death. These signal changes originate from the viscosity contrast between living, substrate-adhering cells, and dead cells which detach from the surface. The sensor directly tracks the detachment process at the surface, enabled by the narrow penetration depth of the acoustic wave (on the order of nanometers).

**Live/dead cell viability assay.** A live/dead fluorescence assay was performed to monitor the effect of a 15% ethanol (EtOH) aqueous solution on MeWo cell viability and adhesion. MeWo cells (ATCC HTB-65) were cultured in Eagle's Minimum Essential Medium supplemented with 10% fetal bovine serum (FBS), 1% penicillin–streptomycin, and 0.05% amphotericin B, without phenol red. The staining solution was freshly prepared before each experiment by diluting Calcein-AM and Ethidium homodimer-1 in phosphate-buffered saline (PBS) at final volumes of 5 and 20 µL, respectively, in 10 mL of PBS. The solution was protected from light during preparation and incubation to prevent fluorophore photobleaching. Cells cultured in 12-well plates were rinsed with PBS, incubated with 1 mL of the staining solution for 30 min at 37 °C in the dark, and subsequently transferred to the fluorescence microscope for imaging. After staining, the PBS containing the fluorescent dyes was replaced by either (i) fresh cell culture medium or (ii) a 15% EtOH aqueous solution. For the control condition, cells were monitored in culture medium for 10 min, with fluorescence images acquired every 60 s. Subsequently, the culture medium was replaced by the 15% EtOH solution, and cell evolution was monitored for an additional 10 min with the same acquisition interval. All experiments were performed under identical imaging conditions.

**Acknowledgments**

JM, FS and MO gratefully acknowledge the support of Centre National d'Études Spatiales (CNES), and La Région Grand Est.
In memory of Sarah Benchabane, who inspired us through her unique outstanding career in this field.